%% file: main.tex
\documentclass[sigconf]{acmart}

\usepackage{booktabs}
\usepackage{graphicx}
\usepackage{enumitem}
\usepackage{amsmath}

\usepackage{amssymb}

\acmConference{}{}{}
\acmYear{}
\acmDOI{}

\setcopyright{none}
\renewcommand\footnotetextcopyrightpermission[1]{}
\title{How Deep Does Your Dependency Tree Go? An Empirical Study of Dependency Amplification Across 10 Package Ecosystems}

\author{Jahidul Arafat}
\email{jza0145@auburn.edu}
\affiliation{%
  \institution{Auburn University}
  \city{Auburn}
  \state{Alabama}
  \country{USA}
}

\begin{document}

\input{sections/abstract}

\maketitle

\input{sections/introduction}
\input{sections/demographics}
\input{sections/rq1}
\input{sections/rq2}
\input{sections/rq3}
\input{sections/discussion}
\input{sections/related_work}
\input{sections/conclusion}

% =============================================================================
% BIBLIOGRAPHY
% =============================================================================

\bibliographystyle{ACM-Reference-Format}
\bibliography{references}

\end{document}

%% file: sections/abstract.tex
\begin{abstract}
Modern software development relies on package ecosystems where declaring a single dependency can introduce dozens of additional transitive packages. This dependency amplification, defined as the ratio of transitive to direct dependencies, has critical implications for software supply chain security, yet no prior work compares amplification patterns across ecosystems at scale. We present an empirical study of 500 projects across 10 major package ecosystems: Maven Central for Java, npm Registry for JavaScript, crates.io for Rust, PyPI for Python, NuGet Gallery for .NET, RubyGems for Ruby, Go Modules for Go, Packagist for PHP, CocoaPods for Swift/Objective-C, and Pub for Dart. Our analysis reveals that Maven exhibits mean amplification of 24.70 times compared to 4.48 times for Go Modules, 4.32 times for npm, and 0.32 times for CocoaPods. We find significant differences with large effect sizes in 22 of 45 pairwise comparisons. These findings challenge prevailing assumptions that npm's preference for small, single-purpose packages leads to the highest amplification. We find that 28\% of Maven projects exhibit amplification exceeding 10 times, indicating elevated amplification is systematic rather than outlier-driven, compared to 14\% for RubyGems, 12\% for npm, and 0\% for Cargo, PyPI, Packagist, CocoaPods, and Pub. We trace these differences to ecosystem design choices including dependency resolution strategies, standard library comprehensiveness, and platform constraints. Our findings suggest practitioners should implement ecosystem-specific security strategies: systematic transitive dependency auditing for Maven environments given 28\% of projects exceed 10 times amplification, targeted outlier identification for npm and RubyGems projects, and continued standard practices for the five ecosystems with controlled amplification. We provide a replication package with data for 500 projects and analysis scripts.
\end{abstract}

%% file: sections/introduction.tex
\section{Introduction}

Modern software development relies on package ecosystems to accelerate development and reduce duplication of effort. Developers declare dependencies on external libraries, leveraging functionality ranging from simple utilities to complex frameworks. However, each declared dependency brings with it a cascade of transitive dependencies. The term transitive dependencies refers to packages required by the direct dependency and its own dependencies, resolved recursively~\cite{decan2019empirical,cox2019surviving,kikas2017structure}. This phenomenon, which we term dependency amplification, has implications for software security and maintenance.

The software supply chain has emerged as a critical attack vector. High-profile incidents such as the event-stream compromise~\cite{ohm2020backstabber}, ua-parser-js malware injection~\cite{ladisa2023sok}, and Log4Shell vulnerability~\cite{wetter2022forensic} demonstrate how vulnerabilities in transitive dependencies can cascade through the software ecosystem. When a developer adds a single dependency, they implicitly trust not only that package but also dozens to hundreds of transitive packages~\cite{zimmermann2019small}, each representing a potential attack surface. Figure~\ref{fig:dependency-example} illustrates how a single Maven project declaring 8 direct dependencies expands to include over 100 additional packages through transitive relationships.

\begin{figure}[t]
\centering
\begin{minipage}{0.95\columnwidth}
\begin{verbatim}
<!-- Maven pom.xml excerpt -->
<dependencies>
  <!-- 8 direct dependencies declared -->
  <dependency>
    <groupId>org.springframework.boot</groupId>
    <artifactId>spring-boot-starter-web</artifactId>
    <version>3.2.0</version>
  </dependency>
  <dependency>
    <groupId>org.springframework.boot</groupId>
    <artifactId>spring-boot-starter-data-jpa</artifactId>
    <version>3.2.0</version>
  </dependency>
  <!-- ... 6 more direct dependencies -->
</dependencies>

<!-- Resolved dependency tree shows 127 total packages -->
<!-- spring-boot-starter-web brings: -->
<!--   spring-web, spring-webmvc, spring-core, -->
<!--   tomcat-embed-core, jackson-databind, ... -->
<!--   (47 transitive packages) -->
<!-- spring-boot-starter-data-jpa brings: -->
<!--   hibernate-core, jakarta.persistence-api, -->
<!--   spring-data-jpa, spring-jdbc, ... -->
<!--   (38 transitive packages) -->
\end{verbatim}
\end{minipage}
\caption{Dependency specification from a typical Maven project. The project directly requires 8 packages, but these bring over 100 additional transitive dependencies. A single Spring Boot dependency transitively depends on dozens of other packages including web servers, JSON processors, and database connectors. Updating any package in this chain can trigger cascading version changes throughout the dependency tree.}
\label{fig:dependency-example}
\end{figure}

Despite growing awareness of supply chain risks, dependency amplification patterns remain uncharacterized across ecosystems using consistent methodology. Conventional wisdom suggests that the npm ecosystem leads to extreme dependency proliferation due to its culture of small, single-purpose packages~\cite{abdalkareem2017why}. However, this assumption lacks empirical validation across multiple ecosystems with consistent methodology and sufficient statistical power. Prior studies have examined individual ecosystems including npm~\cite{zimmermann2019small,chinthanet2021lags}, Maven~\cite{soto2021comprehensive,benelallam2019maven}, PyPI~\cite{alfadel2023empirical}, Cargo~\cite{he2023empirical}, and others, but cross-ecosystem comparison of amplification patterns remains limited. No prior work has examined amplification patterns across 10 or more diverse ecosystems representing different language families, platform targets, and design philosophies.

This paper presents an empirical study of dependency amplification across 10 major package ecosystems representing widely-used programming languages and platforms in modern software development. We analyze Maven Central for Java, npm Registry for JavaScript, crates.io for Rust, PyPI for Python, NuGet Gallery for .NET, RubyGems for Ruby, Go Modules for Go, Packagist for PHP, CocoaPods for Swift/Objective-C, and Pub for Dart. Together they represent diverse design philosophies from enterprise frameworks to platform-constrained development, from dynamic scripting languages to statically-typed systems languages, and from small single-purpose package designs to comprehensive framework architectures.

We answer the following research questions:

\begin{itemize}[leftmargin=*]
\item \textbf{RQ1:} How does dependency amplification vary across 10 major package ecosystems?
\item \textbf{RQ2:} What are the supply chain exposure and propagation scope implications of transitive dependency growth across diverse ecosystems?
\item \textbf{RQ3:} What ecosystem characteristics and design factors explain amplification differences?
\end{itemize}

Our key findings challenge prevailing assumptions about ecosystem amplification profiles and reveal fundamental patterns in dependency management. First, Maven exhibits higher amplification at 24.70 times compared to most ecosystems, with large effect sizes that are significant at the 0.05 level. This contradicts expectations that npm's culture of small, single-purpose packages leads to the highest amplification. Second, amplification patterns diverge across ecosystems where Maven shows 28\% of projects with amplification exceeding 10 times while 5 of 10 ecosystems show 0\% at this threshold. Third, ecosystem design choices impact amplification where platform constraints and standard library comprehensiveness correlate with lower amplification. Fourth, hierarchical clustering reveals Maven occupies an isolated position with extreme amplification while most ecosystems cluster together with controlled amplification below 5 times, suggesting current security frameworks may require ecosystem-specific strategies rather than uniform approaches.

These findings have practical implications for security practitioners and tool developers. Enterprise Java environments may require more aggressive transitive dependency auditing than previously assumed~\cite{plate2015impact,pashchenko2018vulnerable}, while five ecosystems including PyPI, Cargo, and Packagist demonstrate that controlled amplification is achievable through careful design choices~\cite{amann2018study}. The discovery that npm's actual amplification at 4.32 times falls far below Maven's 24.70 times suggests security investment priorities may need substantial revision.

\textbf{Contributions.} We list our contributions as follows:
\begin{itemize}[leftmargin=*]
\item Comprehensive quantification of dependency amplification patterns across 10 major package ecosystems representing diverse language families, platform targets, and design philosophies, with 500 analyzed projects providing robust statistical power.
\item Identification of dramatic ecosystem divergence in supply chain exposure profiles where Maven shows 28\% of projects with amplification exceeding 10 times compared to 0\% for five ecosystems, with hierarchical clustering revealing Maven's isolated position.
\item Discovery that platform constraints and ecosystem design choices correlate with amplification patterns, with Maven at 24.70 times versus CocoaPods at 0.32 times demonstrating a 77-fold difference between highest and lowest amplification ecosystems.
\item Analysis of ecosystem design factors including dependency resolution strategies, zero-dependency prevalence, correlation structures, and variance metrics explaining observed amplification patterns.
\item Complete replication package with dependency data for 500 projects across 10 ecosystems, statistical analysis scripts, and visualization code available for future research.
\end{itemize}

%% file: sections/demographics.tex
\section{Dataset Overview}
\label{sec:dataset}

We provide context on the analyzed package ecosystems in this section. Table~\ref{tab:notation} summarizes the mathematical notation used throughout this paper.

\begin{table}[t]
\centering
\caption{Summary of Notation}
\label{tab:notation}
\small
\begin{tabular}{ll}
\toprule
\textbf{Symbol} & \textbf{Description} \\
\midrule
$p$ & A software project \\
$\mathcal{P}$, $E$ & Set of projects in an ecosystem \\
$D_{direct}(p)$ & Set of directly declared dependencies \\
$D_{transitive}(p)$ & Set of transitively required dependencies \\
$D_{total}(p)$ & Complete dependency set \\
$\alpha(p)$ & Amplification factor \\
$\mathcal{A}(p)$ & Attack surface \\
$\mathcal{I}(v, E)$ & Vulnerability impact function \\
$\mathcal{Z}(p)$ & Zero-dependency indicator \\
$\mathcal{P}(E)$ & Predictability ratio \\
$\rho$ & Spearman's rank correlation \\
$\delta$ & Cliff's delta effect size \\
$CV$ & Coefficient of variation \\
\bottomrule
\end{tabular}
\end{table}

\subsection{Ecosystem Selection}

We selected 10 major package ecosystems representing diverse language families and application domains. Maven Central serves as the primary repository for Java artifacts, supporting enterprise applications, Android development, and backend services. The npm Registry is the largest package ecosystem by package count, serving both frontend and backend JavaScript development. The Cargo registry at crates.io serves the Rust programming community that emphasizes safety and performance. PyPI hosts Python packages dominating data science, machine learning, and scientific computing. NuGet Gallery provides packages for .NET languages across multiple platforms. RubyGems serves the Ruby community. Go Modules manages dependencies for Go emphasizing standard library usage. Packagist provides PHP packages for web development. CocoaPods manages dependencies for Swift and Objective-C in Apple ecosystem development. Pub serves Dart and Flutter development focused on mobile and web UI.

These ecosystems differ in their dependency management approaches. Maven uses XML-based project definitions with explicit version constraints and scope declarations for compile, test, and runtime dependencies. The npm ecosystem uses JSON-based manifest files with semantic versioning ranges and distinguishes between production dependencies, development dependencies, and peer dependencies. Cargo uses TOML-based manifests with strict version resolution and a lock file mechanism, which records exact dependency versions to ensure reproducible builds. Other ecosystems employ distinct manifest formats, version resolution strategies, and dependency scoping mechanisms.

\subsection{Data Collection}

We collected 50 projects from each ecosystem totaling 500 projects. Table~\ref{tab:dataset} summarizes dataset characteristics and technical specifications. Projects were sampled from popular packages to ensure relevance and representativeness.

\begin{table*}[t]
\centering
\caption{Dataset and Ecosystem Characteristics}
\label{tab:dataset}
\resizebox{\textwidth}{!}{%
\begin{tabular}{lrrrrrrrrrr}
\toprule
\textbf{Characteristic} & \textbf{Maven} & \textbf{npm} & \textbf{Cargo} & \textbf{PyPI} & \textbf{NuGet} & \textbf{RubyGems} & \textbf{Go} & \textbf{Packagist} & \textbf{CocoaPods} & \textbf{Pub} \\
\midrule
\multicolumn{11}{l}{\textit{Sample and Dependency Statistics}} \\
Projects analyzed & 50 & 50 & 50 & 50 & 50 & 50 & 50 & 50 & 50 & 50 \\
Mean direct deps & 5.4 & 30.9 & 13.7 & 4.2 & 3.9 & 4.7 & 7.7 & 12.2 & 0.5 & 7.5 \\
Mean transitive deps & 74.9 & 52.8 & 15.1 & 6.1 & 7.7 & 12.1 & 27.3 & 12.2 & 0.7 & 15.8 \\
Mean total deps & 80.3 & 83.7 & 28.8 & 10.2 & 11.6 & 16.8 & 34.9 & 24.5 & 1.2 & 23.2 \\
\midrule
\multicolumn{11}{l}{\textit{Technical Specifications}} \\
Manifest format & XML & JSON & TOML & Python & XML & Ruby & Go & JSON & Ruby & YAML \\
Lock file default & No & Yes & Yes & No & No & Yes & Yes & Yes & Yes & Yes \\
\bottomrule
\end{tabular}%
}
\end{table*}

For Maven, we sampled projects from Maven Central's most-downloaded artifacts including framework components and enterprise utilities. For npm, we sampled packages from the registry's most-depended-upon packages including framework ecosystem components, utility libraries, and build tools. For Cargo, we sampled crates from crates.io's most-downloaded crates including async runtimes, serialization libraries, and web frameworks. For PyPI, we sampled popular packages spanning scientific computing, web frameworks, and machine learning libraries. For other ecosystems, we sampled from frequently-used packages representing diverse application domains.

\subsection{Dependency Resolution}

For each project, we measured two types of dependencies following established terminology in dependency analysis research~\cite{decan2019empirical,kikas2017structure}. Direct dependencies are packages explicitly declared in the project's manifest file. Transitive dependencies are all packages recursively required by direct dependencies, resolved using native package manager tooling to ensure accurate version resolution and conflict handling.

We used each ecosystem's native dependency resolution tooling to ensure measurements reflect actual installed packages. This approach accounts for version resolution, conflict mediation, and platform-specific dependencies. For Maven projects, we extracted complete dependency trees including compile-scope and runtime-scope dependencies. For npm packages, we resolved all production dependencies, development dependencies, and peer dependencies. For Cargo crates, we resolved normal dependencies, development dependencies, and build dependencies. For other ecosystems, we used native tooling to resolve complete dependency trees.

\subsection{Ecosystem Characteristics}

Maven projects declare fewer direct dependencies with mean of 5.4 but accumulate more transitive dependencies. The npm projects declare the most direct dependencies with mean of 30.9 reflecting the ecosystem's composition-focused design. Cargo projects show moderate direct dependency counts with mean of 13.7 and controlled transitive dependency accumulation. PyPI and NuGet show conservative dependency patterns with means of 4.2 and 3.9 direct dependencies respectively. CocoaPods demonstrates minimal external dependency usage with mean of only 0.5 direct dependencies.

These characteristics reflect fundamental design differences. Maven's ecosystem favors comprehensive frameworks providing extensive functionality through single dependencies. The npm ecosystem favors small focused packages following the Unix philosophy of doing one thing well. Cargo's ecosystem emphasizes safety and explicit dependency management with strict version resolution preventing ambiguous package selection. CocoaPods faces platform constraints where most functionality comes from system frameworks rather than external packages.

%% file: sections/rq1.tex
\section{RQ1: Dependency Amplification Across Ecosystems}
\label{sec:rq1}

We provide the methodology and results in Sections~\ref{subsec:rq1-methodology} and~\ref{subsec:rq1-results}.

\subsection{Methodology for RQ1}
\label{subsec:rq1-methodology}

We analyzed 500 projects across 10 ecosystems with 50 projects per ecosystem. For each project, we extracted dependency information from manifest files and resolved complete dependency trees using native package manager tooling.

Our analysis involved three steps. First, we extracted manifest files from each project repository. For Maven projects, we parsed pom.xml files to identify declared dependencies. For npm packages, we parsed package.json files to identify production dependencies, development dependencies, and peer dependencies. For Cargo crates, we parsed Cargo.toml files to identify normal dependencies, development dependencies, and build dependencies. For PyPI packages, we parsed setup.py or pyproject.toml files. For NuGet projects, we parsed .csproj or packages.config files. For RubyGems, we parsed Gemfile specifications. For Go Modules, we parsed go.mod files. For Packagist packages, we parsed composer.json files. For CocoaPods, we parsed Podfile specifications. For Pub packages, we parsed pubspec.yaml files.

Second, we resolved complete dependency trees using native package manager tooling. This step identifies all transitive dependencies by recursively resolving each direct dependency's own requirements. We used Maven's dependency resolution for Java projects, npm's package installation for JavaScript projects, Cargo's dependency resolution for Rust projects, and analogous native tooling for all remaining ecosystems. Using native tooling ensures accurate version resolution accounting for conflict mediation and platform-specific dependencies.

Third, we computed dependency metrics for each project. Let $p$ denote a project and $\mathcal{P}$ denote the set of all projects in an ecosystem. We define the following dependency sets:

\begin{itemize}[leftmargin=*]
\item $D_{direct}(p)$: Set of packages explicitly declared in the manifest file of project $p$
\item $D_{transitive}(p)$: Set of packages recursively required by direct dependencies but not explicitly declared
\item $D_{total}(p)$: Complete dependency footprint where $D_{total}(p) = D_{direct}(p) \cup D_{transitive}(p)$
\end{itemize}

We define the amplification factor $\alpha$ as the ratio of transitive to direct dependencies following prior work on dependency analysis~\cite{zimmermann2019small}:

\begin{equation}
\alpha(p) = \frac{|D_{transitive}(p)|}{\max(|D_{direct}(p)|, 1)}
\label{eq:amplification}
\end{equation}

This metric captures how a single declared dependency expands into multiple installed packages. We use the maximum of direct dependencies and one to handle projects with zero direct dependencies. An amplification factor of $\alpha = 10$ indicates that each direct dependency brings 10 transitive packages.

Given the non-normal distribution of dependency counts confirmed by Shapiro-Wilk tests with all p-values below 0.001, we employ non-parametric statistical methods following established guidelines for software engineering experiments~\cite{arcuri2011practical}.

For omnibus comparison across all 10 ecosystems, we use the Kruskal-Wallis H-test which evaluates whether samples originate from the same distribution:

\begin{equation}
H = \frac{12}{N(N+1)} \sum_{i=1}^{k} \frac{R_i^2}{n_i} - 3(N+1)
\label{eq:kruskal}
\end{equation}

where $N$ is the total sample size, $k$ is the number of groups, $n_i$ is the sample size of group $i$, and $R_i$ is the sum of ranks for group $i$.

For pairwise ecosystem comparisons, we use Mann-Whitney U tests with Holm-Bonferroni correction for multiple comparisons. With 10 ecosystems we perform 45 pairwise comparisons. We compute Cliff's delta $\delta$ for effect size estimation~\cite{cliff1993dominance}:

\begin{equation}
\delta = \frac{|\{(x_i, y_j) : x_i > y_j\}| - |\{(x_i, y_j) : x_i < y_j\}|}{n_1 \cdot n_2}
\label{eq:cliff}
\end{equation}

where $x_i$ are observations from group 1, $y_j$ are observations from group 2, and $n_1$, $n_2$ are respective sample sizes. We interpret effect sizes following Romano et al~\cite{romano2006appropriate}: $|\delta| < 0.147$ as negligible, $|\delta| < 0.33$ as small, $|\delta| < 0.474$ as medium, and $|\delta| \geq 0.474$ as large. We report 95\% confidence intervals computed via bootstrap resampling with 10,000 iterations.

\subsection{Answer to RQ1}
\label{subsec:rq1-results}

Table~\ref{tab:rq1_descriptive} presents descriptive statistics for dependency counts and amplification factors across ecosystems.

\begin{table*}[t]
\centering
\caption{Dependency Characteristics Across 10 Ecosystems}
\label{tab:rq1_descriptive}
\small
\begin{tabular}{lrrrr}
\toprule
\textbf{Ecosystem} & \textbf{Direct} & \textbf{Transitive} & \textbf{Total} & \textbf{Amplification} \\
\midrule
Maven & 5.4 $\pm$ 6.8 & 74.9 $\pm$ 133.7 & 80.3 $\pm$ 133.7 & 24.70$\times$ \\
Go Modules & 7.7 $\pm$ 13.7 & 27.3 $\pm$ 41.8 & 34.9 $\pm$ 49.3 & 4.48$\times$ \\
npm & 30.9 $\pm$ 39.5 & 52.8 $\pm$ 114.0 & 83.7 $\pm$ 134.6 & 4.32$\times$ \\
RubyGems & 4.7 $\pm$ 5.2 & 12.1 $\pm$ 15.2 & 16.8 $\pm$ 17.7 & 4.32$\times$ \\
Pub & 7.5 $\pm$ 5.7 & 15.8 $\pm$ 13.3 & 23.2 $\pm$ 16.2 & 2.61$\times$ \\
NuGet & 3.9 $\pm$ 7.3 & 7.7 $\pm$ 21.4 & 11.6 $\pm$ 26.6 & 2.32$\times$ \\
PyPI & 4.2 $\pm$ 5.2 & 6.1 $\pm$ 7.2 & 10.2 $\pm$ 10.3 & 1.50$\times$ \\
Packagist & 12.2 $\pm$ 11.6 & 12.2 $\pm$ 15.5 & 24.5 $\pm$ 24.0 & 1.24$\times$ \\
Cargo & 13.7 $\pm$ 15.0 & 15.1 $\pm$ 29.7 & 28.8 $\pm$ 40.5 & 0.97$\times$ \\
CocoaPods & 0.5 $\pm$ 1.1 & 0.7 $\pm$ 1.4 & 1.2 $\pm$ 2.1 & 0.32$\times$ \\
\bottomrule
\multicolumn{5}{l}{\footnotesize{Values shown as mean $\pm$ standard deviation.}}
\end{tabular}
\end{table*}

Maven exhibits the highest amplification with mean of 24.70 times, indicating that a Maven project declaring 5.4 direct dependencies installs 74.9 transitive packages. This finding contradicts conventional expectations that npm's preference for small, single-purpose packages leads to the highest amplification. The npm ecosystem shows mean amplification of 4.32 times, tied with RubyGems at 4.32 times. Go Modules shows amplification of 4.48 times despite its minimalist standard library philosophy. Pub demonstrates moderate amplification at 2.61 times while NuGet shows 2.32 times.

PyPI exhibits controlled amplification at 1.50 times, Packagist shows 1.24 times, and Cargo demonstrates amplification below unity at 0.97 times indicating transitive dependencies barely exceed direct declarations. CocoaPods shows the lowest amplification at only 0.32 times, reflecting platform constraints where most functionality comes from system frameworks.

The Kruskal-Wallis H-test confirms significant differences across ecosystems with H statistic of 150.90 and p-value of $5.74 \times 10^{-28}$. Table~\ref{tab:rq1_pairwise} presents pairwise comparisons with effect sizes.

\begin{table}[t]
\centering
\caption{Pairwise Ecosystem Comparisons for Amplification}
\label{tab:rq1_pairwise}
\small
\begin{tabular}{lrrr}
\toprule
\textbf{Comparison} & \textbf{p-value} & \textbf{Cliff's $\delta$} & \textbf{Effect} \\
\midrule
Go Modules vs CocoaPods & $1.14 \times 10^{-16}$ & 0.932 & large \\
CocoaPods vs Pub & $6.26 \times 10^{-14}$ & -0.848 & large \\
Cargo vs Go Modules & $2.83 \times 10^{-13}$ & -0.846 & large \\
Maven vs CocoaPods & $2.92 \times 10^{-11}$ & 0.750 & large \\
Maven vs Cargo & $2.02 \times 10^{-8}$ & 0.634 & large \\
Maven vs npm & $1.85 \times 10^{-6}$ & 0.494 & large \\
Maven vs PyPI & $8.61 \times 10^{-6}$ & 0.465 & medium \\
npm vs Cargo & $9.85 \times 10^{-5}$ & 0.394 & small \\
\bottomrule
\multicolumn{4}{l}{\footnotesize Significant after Holm-Bonferroni correction}
\end{tabular}
\end{table}

Maven differs from all other ecosystems with large effect sizes for comparisons against CocoaPods with Cliff's delta of 0.750, against Cargo with delta of 0.634, and against npm with delta of 0.494. The comparison between Maven and npm represents the contrast between the two ecosystems often considered to have highest amplification. Maven's higher amplification challenges assumptions about npm being the primary concern.

\begin{figure}[t]
\centering
\includegraphics[width=0.48\textwidth]{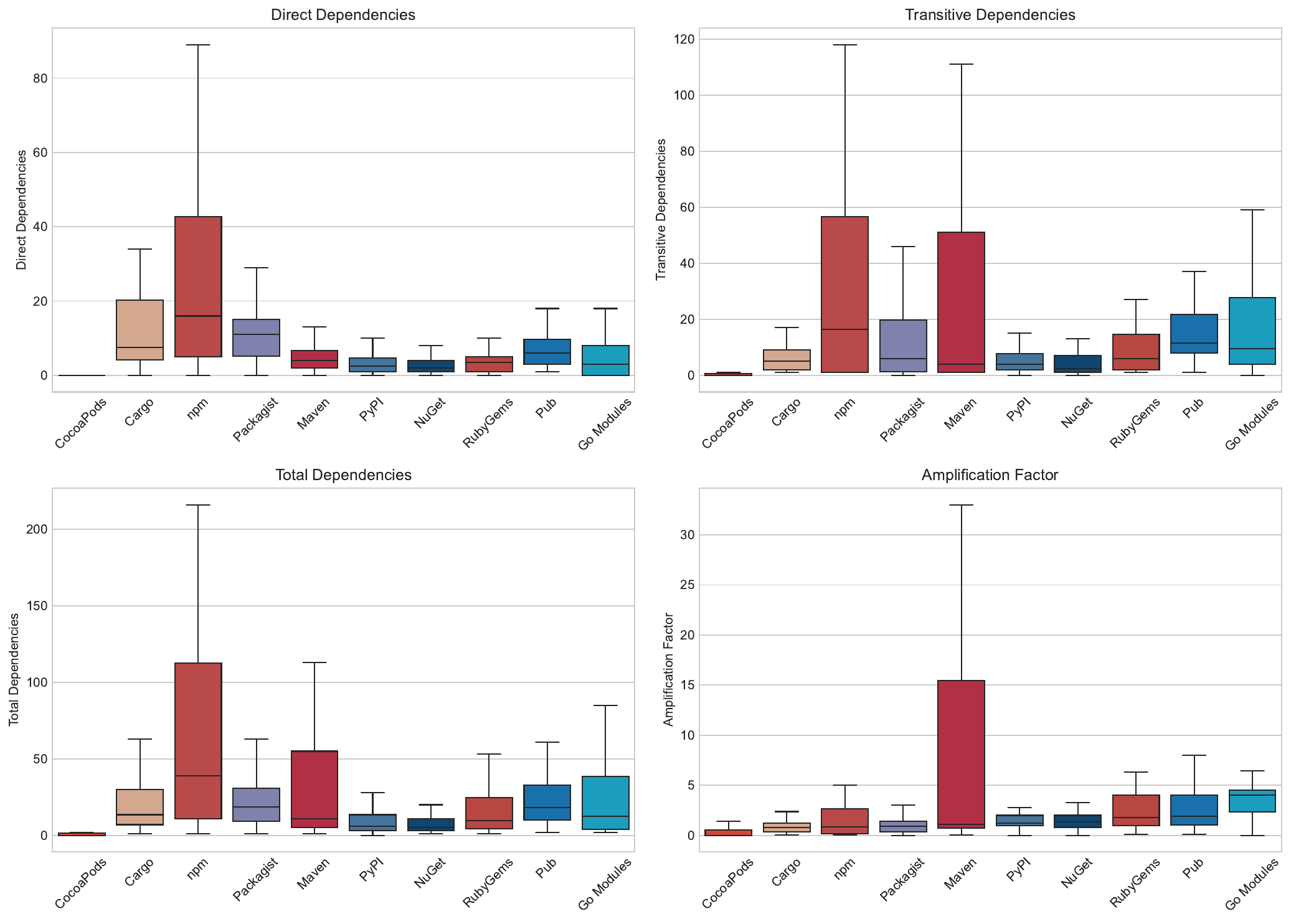}
\caption{Dependency amplification distribution across 10 ecosystems. Maven exhibits significantly higher amplification with extreme outliers reaching 198.5 times. Cargo and CocoaPods maintain controlled amplification below 4 times.}
\label{fig:rq1_boxplot}
\end{figure}

Figure~\ref{fig:rq1_boxplot} visualizes amplification distributions. Maven shows not only higher median amplification but also extreme outliers with maximum amplification reaching 198.5 times. The 95th percentile for Maven at 112.7 times far exceeds Go Modules at 13.9 times, npm at 22.1 times, and all remaining ecosystems.

Distribution shapes differ across ecosystems. Maven's amplification distribution shows high positive skewness indicating a long right tail with extreme values. The npm and Go Modules ecosystems show even higher variance reflecting heterogeneous package composition. Cargo, PyPI, Packagist, and Pub demonstrate more controlled distributions. CocoaPods shows the most constrained distribution with maximum amplification of only 2.0 times.

Table~\ref{tab:rq1_extended} presents extended statistics with confidence intervals and percentile information.

\begin{table}[t]
\centering
\caption{Extended Amplification Statistics}
\label{tab:rq1_extended}
\small
\begin{tabular}{lrrrrr}
\toprule
\textbf{Ecosystem} & \textbf{Mean} & \textbf{95\% CI} & \textbf{Median} & \textbf{P95} & \textbf{Max} \\
\midrule
Maven & 24.70 & [11.8, 40.9] & 4.00 & 112.7 & 198.5 \\
Go Modules & 4.48 & [2.5, 6.8] & 1.67 & 13.9 & 22.0 \\
npm & 4.32 & [1.5, 8.5] & 0.96 & 22.1 & 76.2 \\
RubyGems & 4.32 & [2.5, 6.5] & 2.00 & 11.9 & 50.0 \\
Pub & 2.61 & [2.0, 3.3] & 2.14 & 5.1 & 8.5 \\
NuGet & 2.32 & [0.9, 4.3] & 0.67 & 7.7 & 23.0 \\
PyPI & 1.50 & [1.1, 2.0] & 1.00 & 2.9 & 5.0 \\
Packagist & 1.24 & [0.9, 1.6] & 0.92 & 2.3 & 5.3 \\
Cargo & 0.97 & [0.7, 1.3] & 0.75 & 2.3 & 3.9 \\
CocoaPods & 0.32 & [0.1, 0.5] & 0.00 & 1.3 & 2.0 \\
\bottomrule
\end{tabular}
\end{table}

The wide confidence intervals for Maven and npm reflect high variance in these ecosystems. Maven's 95\% confidence interval spans from 11.8 to 40.9 times. CocoaPods shows a narrow interval from 0.1 to 0.5 times indicating consistent low amplification across projects. PyPI and Packagist similarly show narrow confidence intervals reflecting consistent behavior.

\textbf{Answer to RQ1:} Maven exhibits higher dependency amplification at 24.70 times compared to Go Modules at 4.48 times, npm at 4.32 times, RubyGems at 4.32 times, and Cargo at 0.97 times. CocoaPods shows the lowest amplification at 0.32 times. These differences are significant at the 0.05 level with large effect sizes in 22 of 45 pairwise comparisons. Maven's amplification reaches extreme values up to 198.5 times while Cargo and CocoaPods maintain controlled amplification below 4 times. These findings challenge assumptions that npm's culture of small, single-purpose packages leads to the highest amplification.

%% file: sections/rq2.tex
\section{RQ2: Supply Chain Exposure and Propagation Scope}
\label{sec:rq2}

We provide the methodology and results in Sections~\ref{subsec:rq2-methodology} and~\ref{subsec:rq2-results}.

\subsection{Methodology for RQ2}
\label{subsec:rq2-methodology}

We analyze supply chain exposure and propagation scope using formal metrics for attack surface, concentration ratios, and vulnerability propagation potential. While we do not analyze specific CVEs, these metrics quantify the extent of code trust and the potential scope of impact if vulnerabilities occur.

We define the attack surface $\mathcal{A}$ of a project $p$ as the total number of distinct packages that must be trusted:

\begin{equation}
\mathcal{A}(p) = |D_{total}(p)| = |D_{direct}(p)| + |D_{transitive}(p)|
\label{eq:attack_surface}
\end{equation}

Each package in $D_{total}(p)$ represents code that executes within the application context, creating potential sources of vulnerabilities, malicious code injection, or maintenance abandonment.

We analyze amplification distribution by comparing mean and maximum attack surfaces to distinguish consistent versus outlier-driven patterns. Let $\mu_{\mathcal{A}}(E)$ denote mean attack surface and $\max_{\mathcal{A}}(E)$ denote maximum attack surface for ecosystem $E$. We define the concentration ratio:

\begin{equation}
\mathcal{C}(E) = \frac{\max_{\mathcal{A}}(E)}{\mu_{\mathcal{A}}(E)}
\label{eq:concentration}
\end{equation}

High concentration ratio indicates amplification concentrated in outliers while low ratio indicates consistent amplification across projects.

We model vulnerability propagation potential using the impact function $\mathcal{I}$. When a vulnerability occurs in package $v$, the number of affected projects depends on both direct and transitive usage:

\begin{equation}
\mathcal{I}(v, E) = \sum_{p \in E} \mathbb{1}[v \in D_{total}(p)]
\label{eq:impact}
\end{equation}

where $\mathbb{1}[\cdot]$ is the indicator function. We estimate expected propagation by multiplying direct dependents by mean amplification: if package $v$ has $d$ direct dependents, the estimated total affected projects is $d \cdot \bar{\alpha}(E)$ where $\bar{\alpha}(E)$ is the mean amplification factor for ecosystem $E$.

For statistical analysis of exposure metrics, we use non-parametric tests given the non-normal distributions. We compute percentiles at 90th, 95th, and 99th levels to characterize distribution tails where supply chain risks concentrate. We report bootstrap confidence intervals for prevalence estimates.

To visualize amplification distribution patterns, we compute cumulative distribution functions (CDFs) for amplification factors. The CDF $F(x)$ represents the proportion of projects with amplification factor less than or equal to $x$:

\begin{equation}
F(x) = P(\alpha(p) \leq x) = \frac{|\{p \in E : \alpha(p) \leq x\}|}{|E|}
\label{eq:cdf}
\end{equation}

Steep CDF curves indicate consistent low amplification while gradual curves with long tails indicate substantial proportions of high-amplification projects.

\subsection{Answer to RQ2}
\label{subsec:rq2-results}

Table~\ref{tab:rq2_security} presents attack surface and supply chain exposure metrics across ecosystems.

\begin{table}[t]
\centering
\caption{Attack Surface and Supply Chain Exposure Analysis}
\label{tab:rq2_security}
\small
\begin{tabular}{lrrrr}
\toprule
\textbf{Ecosystem} & \textbf{Mean} & \textbf{Max} & \textbf{P95} & \textbf{$\alpha \geq 10$} \\
\midrule
npm & 83.7 & 785 & 250.6 & 6 (12\%) \\
Maven & 80.3 & 450 & 397.2 & 14 (28\%) \\
Go Modules & 34.9 & 263 & 133.6 & 3 (6\%) \\
Cargo & 28.8 & 189 & 132.4 & 0 (0\%) \\
Packagist & 24.5 & 132 & 63.0 & 0 (0\%) \\
Pub & 23.2 & 61 & 54.3 & 0 (0\%) \\
RubyGems & 16.8 & 78 & 49.6 & 7 (14\%) \\
NuGet & 11.6 & 163 & 30.0 & 2 (4\%) \\
PyPI & 10.2 & 57 & 30.2 & 0 (0\%) \\
CocoaPods & 1.2 & 12 & 7.0 & 0 (0\%) \\
\bottomrule
\multicolumn{5}{l}{\footnotesize{Attack surface measured as total packages installed.}} \\
\multicolumn{5}{l}{\footnotesize{$\alpha \geq 10$: Projects with amplification factor exceeding 10 times.}}
\end{tabular}
\end{table}

Maven shows consistently elevated attack surfaces with mean of 80.3 packages. This means the average Maven project trusts code from over 80 distinct packages, each representing potential sources of vulnerabilities, malicious code, or maintenance issues. The npm ecosystem shows similar mean attack surface at 83.7 packages but higher maximum at 785 packages indicating extreme outliers.

The concentration ratio $\mathcal{C}$ quantifies whether amplification is systematic or outlier-driven. The npm ecosystem shows concentration ratio of $\mathcal{C}_{npm} = 9.4$ indicating amplification concentrated in extreme outliers. Maven shows concentration ratio of $\mathcal{C}_{maven} = 5.6$ indicating more systematic amplification distribution. Cargo shows $\mathcal{C}_{cargo} = 6.6$ while CocoaPods shows $\mathcal{C}_{cocoapods} = 10.0$, both reflecting occasional outliers in otherwise controlled ecosystems.

\begin{figure}[t]
\centering
\includegraphics[width=0.48\textwidth]{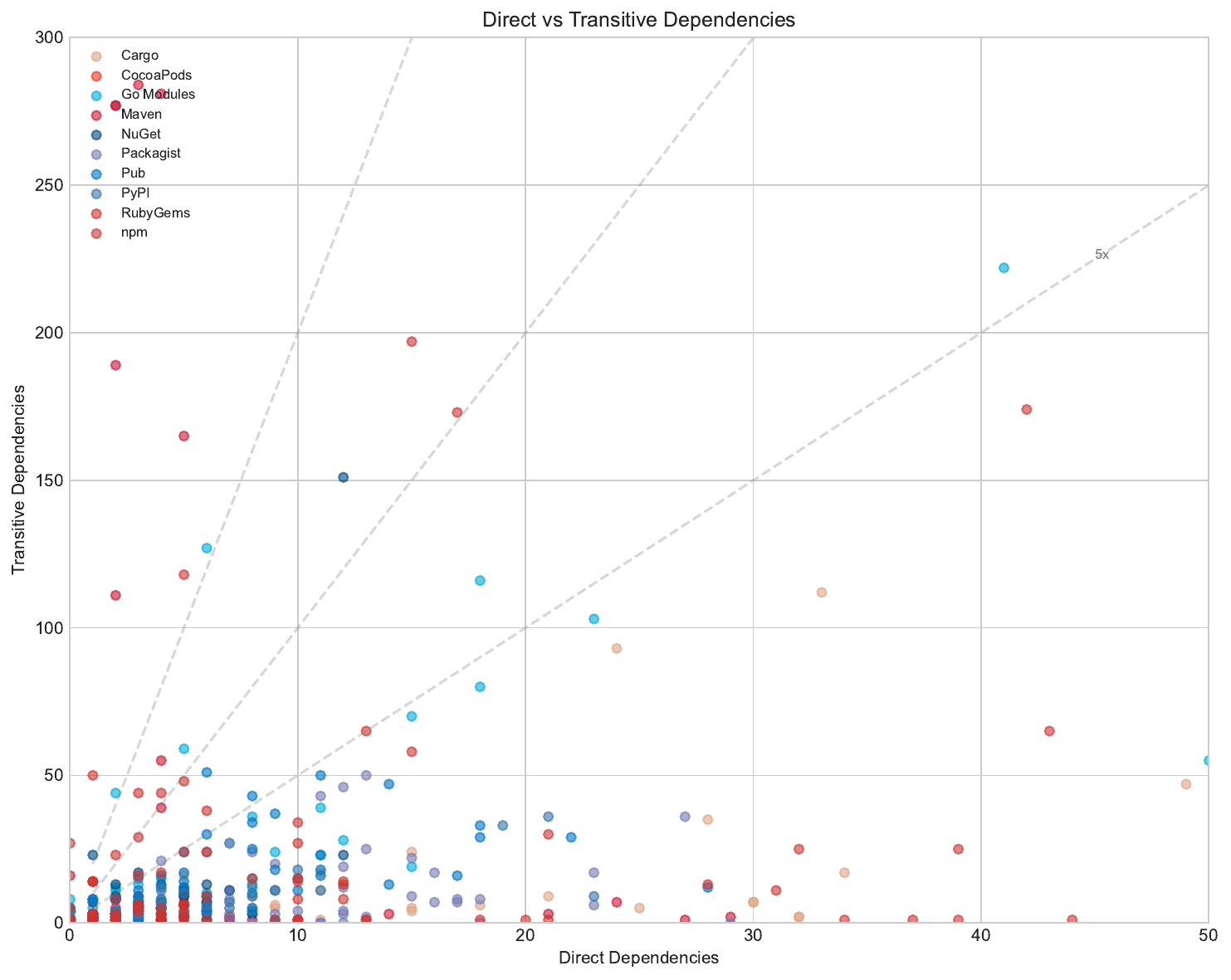}
\caption{Direct versus total dependencies across 10 ecosystems. Maven and RubyGems show steep amplification slopes indicating systematic elevated amplification. The npm ecosystem shows higher variance with some projects reaching extreme values while others maintain moderate totals. Cargo, PyPI, and Packagist maintain controlled linear growth across all projects.}
\label{fig:rq2_scatter}
\end{figure}

Maven exhibits 28\% of projects with amplification exceeding 10 times compared to 14\% for RubyGems, 12\% for npm, and 0\% for five ecosystems including Cargo, PyPI, Packagist, CocoaPods, and Pub. This finding indicates Maven's elevated amplification is systematic rather than concentrated in outliers. Figure~\ref{fig:rq2_scatter} visualizes this pattern showing Maven's steep amplification slope.

The npm ecosystem shows the largest maximum attack surface at 785 packages from a single project, exceeding Maven's maximum of 450 packages. However, npm's mean attack surface remains comparable to Maven indicating that extreme cases are outliers rather than typical behavior. This distinction has practical implications: npm's elevated amplification can be managed by identifying specific high-amplification packages while Maven requires systematic auditing across most projects.

Cargo and five other ecosystems maintain controlled attack surfaces with 0\% of projects exceeding 10 times amplification. Even at the 95th percentile, Cargo's attack surface of 132.4 packages and PyPI's of 30.2 packages remain moderate compared to Maven's 397.2 packages. This controlled behavior suggests these ecosystems' designs limit transitive dependency growth.

\begin{figure}[t]
\centering
\includegraphics[width=0.48\textwidth]{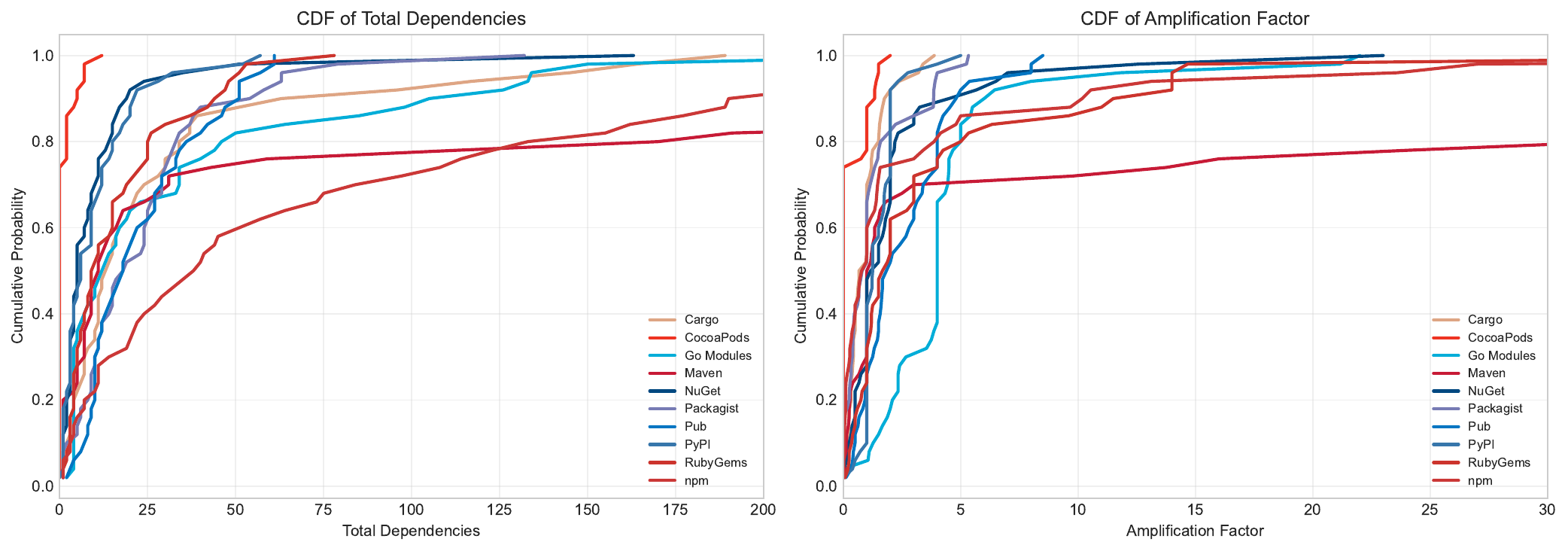}
\caption{Cumulative distribution functions of amplification factor across 10 ecosystems. Cargo and CocoaPods show steep rise indicating consistent low amplification. Maven and RubyGems show gradual rise with long tail. The npm ecosystem shows bimodal pattern with concentration at low amplification but extreme outliers.}
\label{fig:rq2_cdf}
\end{figure}

Figure~\ref{fig:rq2_cdf} shows cumulative distribution functions of amplification factors. CocoaPods and Cargo show steep rise indicating consistent low amplification across projects. Maven shows gradual rise with long tail indicating substantial portion of projects with high amplification. The npm ecosystem shows intermediate pattern with most projects below 10 times amplification but extreme outliers.

\textbf{Answer to RQ2:} Maven exhibits elevated supply chain exposure with 28\% of projects showing amplification exceeding 10 times, compared to 14\% for RubyGems, 12\% for npm, and 0\% for five ecosystems. While npm shows the largest maximum attack surface at 785 packages, its elevated amplification concentrates in outliers rather than typical projects. Propagation scope is highest in Maven where a single vulnerable package can affect over 24 times more projects than direct dependency counts suggest. Cargo, PyPI, Packagist, CocoaPods, and Pub maintain controlled attack surfaces with no projects exceeding 10 times amplification.

%% file: sections/rq3.tex
\section{RQ3: Ecosystem Design Factors Explaining Amplification Differences}
\label{sec:rq3}

We provide the methodology and results in Sections~\ref{subsec:rq3-methodology} and~\ref{subsec:rq3-results}.

\subsection{Methodology for RQ3}
\label{subsec:rq3-methodology}

We investigated ecosystem characteristics that may explain the observed amplification differences through three analytical approaches: zero-dependency analysis, correlation analysis, and ecosystem design examination.

For zero-dependency analysis, we define the zero-dependency indicator function:

\begin{equation}
\mathcal{Z}(p) = \mathbb{1}[|D_{direct}(p)| = 0]
\label{eq:zerodep}
\end{equation}

Zero-dependency packages represent self-contained functionality with no external requirements. We compute zero-dependency prevalence for ecosystem $E$ as:

\begin{equation}
\mathcal{Z}_E = \frac{\sum_{p \in E} \mathcal{Z}(p)}{|E|}
\label{eq:zerodep_prevalence}
\end{equation}

We report bootstrap 95\% confidence intervals for prevalence estimates and test for ecosystem differences using chi-square test for independence.

For correlation analysis, we compute Spearman's rank correlation coefficient $\rho$ between direct and transitive dependency counts:

\begin{equation}
\rho = 1 - \frac{6 \sum_{i=1}^{n} d_i^2}{n(n^2 - 1)}
\label{eq:spearman}
\end{equation}

where $d_i$ is the difference between ranks of paired observations and $n$ is sample size. Strong positive correlation indicates predictable amplification where direct dependency counts reliably predict transitive growth. Weak correlation indicates unpredictable amplification where direct counts provide limited information about total footprint.

We define the predictability ratio $\mathcal{P}$ as the correlation between direct and total dependencies:

\begin{equation}
\mathcal{P}(E) = \rho(|D_{direct}|, |D_{total}|) \text{ for } p \in E
\label{eq:predictability}
\end{equation}

High predictability ratio near 1.0 indicates developers can estimate supply chain exposure from direct dependency counts alone.

For variance analysis, we compute coefficient of variation to measure consistency within ecosystems:

\begin{equation}
CV(E) = \frac{\sigma_\alpha(E)}{\mu_\alpha(E)}
\label{eq:cv}
\end{equation}

where $\sigma_\alpha(E)$ and $\mu_\alpha(E)$ are standard deviation and mean of amplification factors in ecosystem $E$. High $CV$ indicates heterogeneous amplification patterns while low $CV$ indicates consistent behavior across projects.

We also compute the Gini coefficient $G$ to measure inequality in total dependency distribution within each ecosystem. The Gini coefficient ranges from 0 (perfect equality where all projects have identical dependency counts) to 1 (perfect inequality where one project has all dependencies):

\begin{equation}
G(E) = \frac{\sum_{i=1}^{n} \sum_{j=1}^{n} |D_{total}(p_i) - D_{total}(p_j)|}{2n^2\bar{D}}
\label{eq:gini}
\end{equation}

where $n = |E|$ is the number of projects in ecosystem $E$ and $\bar{D}$ is the mean total dependency count. High Gini coefficient indicates a few projects dominate the dependency landscape while most remain lightweight.

For ecosystem design analysis, we examine how package manager architecture influences amplification. We analyze version resolution strategies, dependency scope distinctions, and lock file practices across ecosystems. We also perform hierarchical clustering using Ward's linkage method with Euclidean distance on standardized amplification metrics to identify natural ecosystem groupings.

\subsection{Answer to RQ3}
\label{subsec:rq3-results}

Table~\ref{tab:rq3_zerodeps} presents zero-dependency analysis results.

\begin{table}[t]
\centering
\caption{Zero-Dependency Package Analysis}
\label{tab:rq3_zerodeps}
\small
\begin{tabular}{l@{\hspace{1em}}r@{\hspace{1em}}r@{\hspace{1em}}r}
\toprule
\textbf{Ecosystem} & \textbf{Count} & \textbf{\%} & \textbf{95\% CI} \\
\midrule
CocoaPods & 38/50 & 76.0 & [63.2, 88.8] \\
npm & 20/50 & 40.0 & [26.4, 53.6] \\
Go Modules & 18/50 & 36.0 & [22.6, 49.4] \\
PyPI & 10/50 & 20.0 & [9.0, 31.0] \\
Pub & 10/50 & 20.0 & [9.0, 31.0] \\
RubyGems & 8/50 & 16.0 & [6.0, 26.0] \\
NuGet & 6/50 & 12.0 & [3.4, 20.6] \\
Cargo & 5/50 & 10.0 & [2.0, 18.0] \\
Packagist & 5/50 & 10.0 & [2.0, 18.0] \\
Maven & 14/50 & 28.0 & [15.8, 40.2] \\
\bottomrule
\end{tabular}
\end{table}

CocoaPods shows the highest proportion of zero-dependency packages at $\mathcal{Z}_{cocoapods} = 76.0\%$ reflecting platform constraints where most functionality comes from Apple system frameworks rather than external packages. The npm ecosystem shows $\mathcal{Z}_{npm} = 40.0\%$ and Go Modules shows $\mathcal{Z}_{go} = 36.0\%$ consistent with preferences for small, single-purpose utilities that often have no dependencies. Maven shows moderate zero-dependency prevalence at $\mathcal{Z}_{maven} = 28.0\%$ while Cargo and Packagist show lowest at $\mathcal{Z}_{cargo} = \mathcal{Z}_{packagist} = 10.0\%$. The chi-square test for ecosystem differences is significant with $\chi^2 = 91.4$ and $p < 0.001$.

Zero-dependency packages contribute to ecosystem amplification patterns. When a package has zero direct dependencies, its amplification is zero by our formula. This partially explains why npm's average of 4.32 times remains lower than Maven's 24.70 times despite npm's reputation for deep dependency trees. However, zero-dependency prevalence alone cannot explain all differences as Maven shows 28.0\% zero-dependency packages yet maintains highest amplification.

Table~\ref{tab:rq3_correlation} presents correlation analysis results.

\begin{table}[t]
\centering
\caption{Correlation Between Direct and Transitive Dependencies}
\label{tab:rq3_correlation}
\small
\begin{tabular}{l@{\hspace{1em}}r@{\hspace{1em}}r}
\toprule
\textbf{Ecosystem} & \textbf{$\rho$} & \textbf{p-value} \\
\midrule
CocoaPods & 0.950 & $6.56 \times 10^{-26}$ \\
PyPI & 0.906 & $1.54 \times 10^{-19}$ \\
Go Modules & 0.838 & $3.01 \times 10^{-14}$ \\
Cargo & 0.681 & $5.28 \times 10^{-8}$ \\
Pub & 0.598 & $4.57 \times 10^{-6}$ \\
npm & 0.398 & 0.004 \\
NuGet & 0.397 & 0.004 \\
RubyGems & 0.368 & 0.008 \\
Packagist & 0.342 & 0.015 \\
Maven & 0.196 & 0.172 \\
\bottomrule
\end{tabular}
\end{table}

Maven shows the weakest correlation between direct and transitive dependencies with $\rho_{maven} = 0.196$ which is not significant at the 0.05 level. This weak correlation indicates Maven's transitive dependency growth is unpredictable from direct dependency counts. Adding one dependency to a Maven project may introduce varying numbers of transitive packages depending on which framework components are included.

CocoaPods shows the strongest correlation with $\rho_{cocoapods} = 0.950$ indicating predictable amplification behavior. PyPI shows $\rho_{pypi} = 0.906$ and Go Modules shows $\rho_{go} = 0.838$, both indicating strong predictability. Cargo shows $\rho_{cargo} = 0.681$ indicating moderate-to-strong predictability. These ecosystems demonstrate that predictable amplification is achievable through careful design.

We also examined the predictability ratio $\mathcal{P}$ measuring correlation between direct and total dependencies. CocoaPods shows near-perfect predictability with $\mathcal{P}_{cocoapods} = 0.989$ indicating total dependency footprint is highly predictable from direct dependencies. PyPI shows $\mathcal{P}_{pypi} = 0.947$ and Cargo shows $\mathcal{P}_{cargo} = 0.894$. Maven's weak predictability at $\mathcal{P}_{maven} = 0.433$ means direct dependency counts provide limited information about actual supply chain exposure.

Table~\ref{tab:rq3_variance} presents variance analysis showing amplification consistency within ecosystems.

\begin{table}[t]
\centering
\caption{Amplification Variance by Ecosystem}
\label{tab:rq3_variance}
\small
\begin{tabular}{l@{\hspace{1em}}r@{\hspace{1em}}r@{\hspace{1em}}r}
\toprule
\textbf{Ecosystem} & \textbf{Mean} & \textbf{Std Dev} & \textbf{CV} \\
\midrule
npm & 4.32 & 11.74 & 272\% \\
Maven & 24.70 & 60.56 & 245\% \\
RubyGems & 4.32 & 9.37 & 217\% \\
NuGet & 2.32 & 5.31 & 229\% \\
CocoaPods & 0.32 & 0.55 & 173\% \\
Cargo & 0.97 & 1.37 & 141\% \\
Go Modules & 4.48 & 6.29 & 140\% \\
PyPI & 1.50 & 1.52 & 101\% \\
Packagist & 1.24 & 1.22 & 98\% \\
Pub & 2.61 & 1.83 & 70\% \\
\bottomrule
\multicolumn{4}{l}{\footnotesize{CV (Coefficient of Variation) = (Std Dev / Mean) $\times$ 100\%.}}
\end{tabular}
\end{table}

The npm ecosystem shows highest coefficient of variation at $CV_{npm} = 272\%$ reflecting extreme heterogeneity where amplification ranges from zero to 76.2 times. Maven shows $CV_{maven} = 245\%$ and RubyGems shows $CV_{rubygems} = 217\%$ indicating high variability. Pub shows the lowest coefficient of variation at $CV_{pub} = 70\%$ followed by Packagist at $CV_{packagist} = 98\%$ and PyPI at $CV_{pypi} = 101\%$ indicating more consistent amplification patterns across projects.

\begin{figure}[t]
\centering
\includegraphics[width=0.48\textwidth]{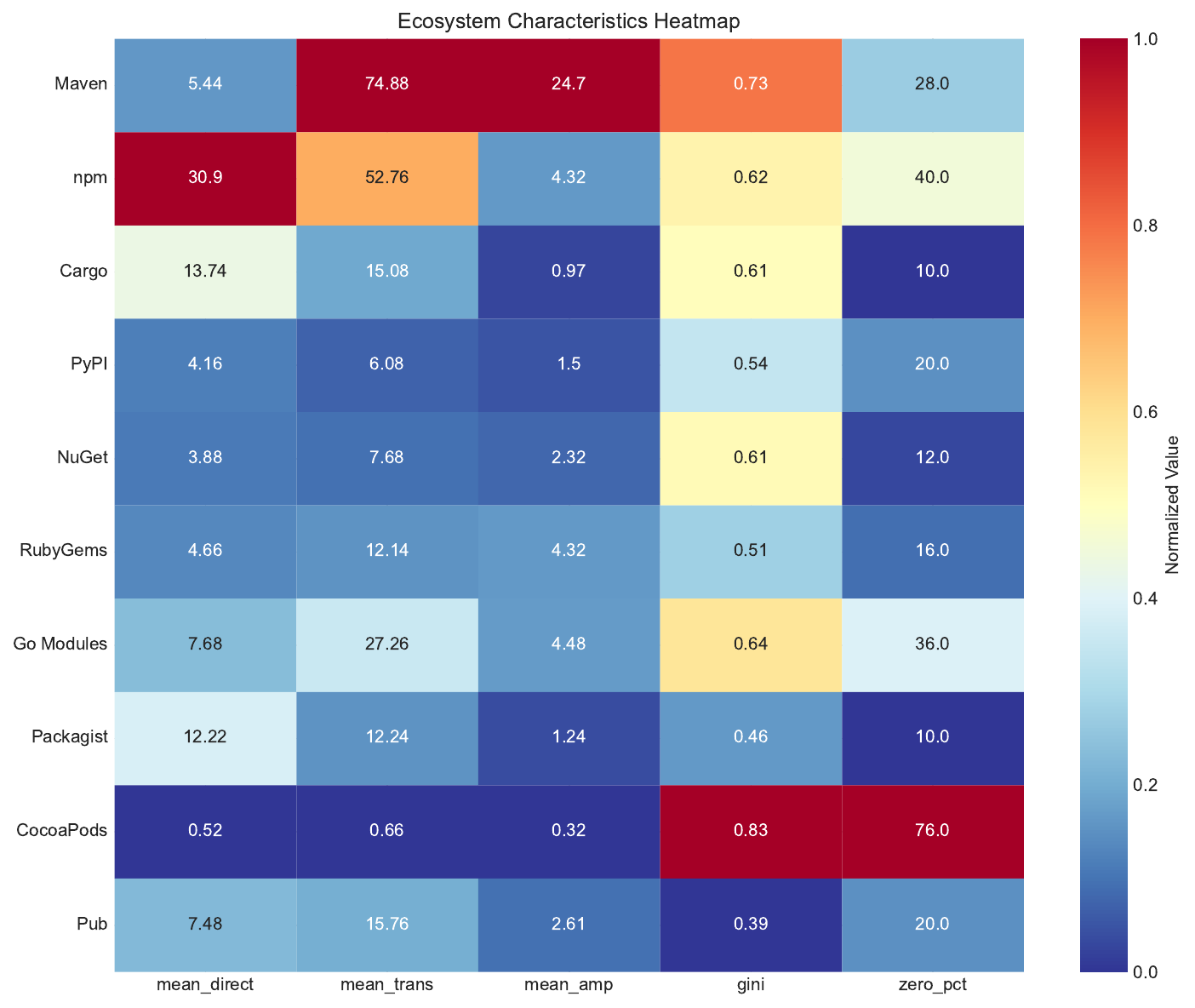}
\caption{Heatmap of normalized ecosystem characteristics across five metrics: mean direct dependencies, mean transitive dependencies, mean amplification, Gini coefficient, and zero-dependency percentage. Darker colors indicate higher normalized values. Maven shows extreme amplification and transitive dependency characteristics. CocoaPods demonstrates minimal dependency footprint. Color intensity reveals natural ecosystem clustering with Maven isolated and most ecosystems forming a controlled-amplification cluster.}
\label{fig:rq3_heatmap}
\end{figure}

Figure~\ref{fig:rq3_heatmap} presents a heatmap visualization of ecosystem characteristics showing normalized values for key dependency metrics. The heatmap reveals clear clustering where Maven exhibits extreme values for transitive dependencies and amplification while CocoaPods shows minimal external dependency usage. PyPI, Packagist, and Cargo form a moderate cluster with balanced characteristics.

\begin{figure}[t]
\centering
\includegraphics[width=0.48\textwidth]{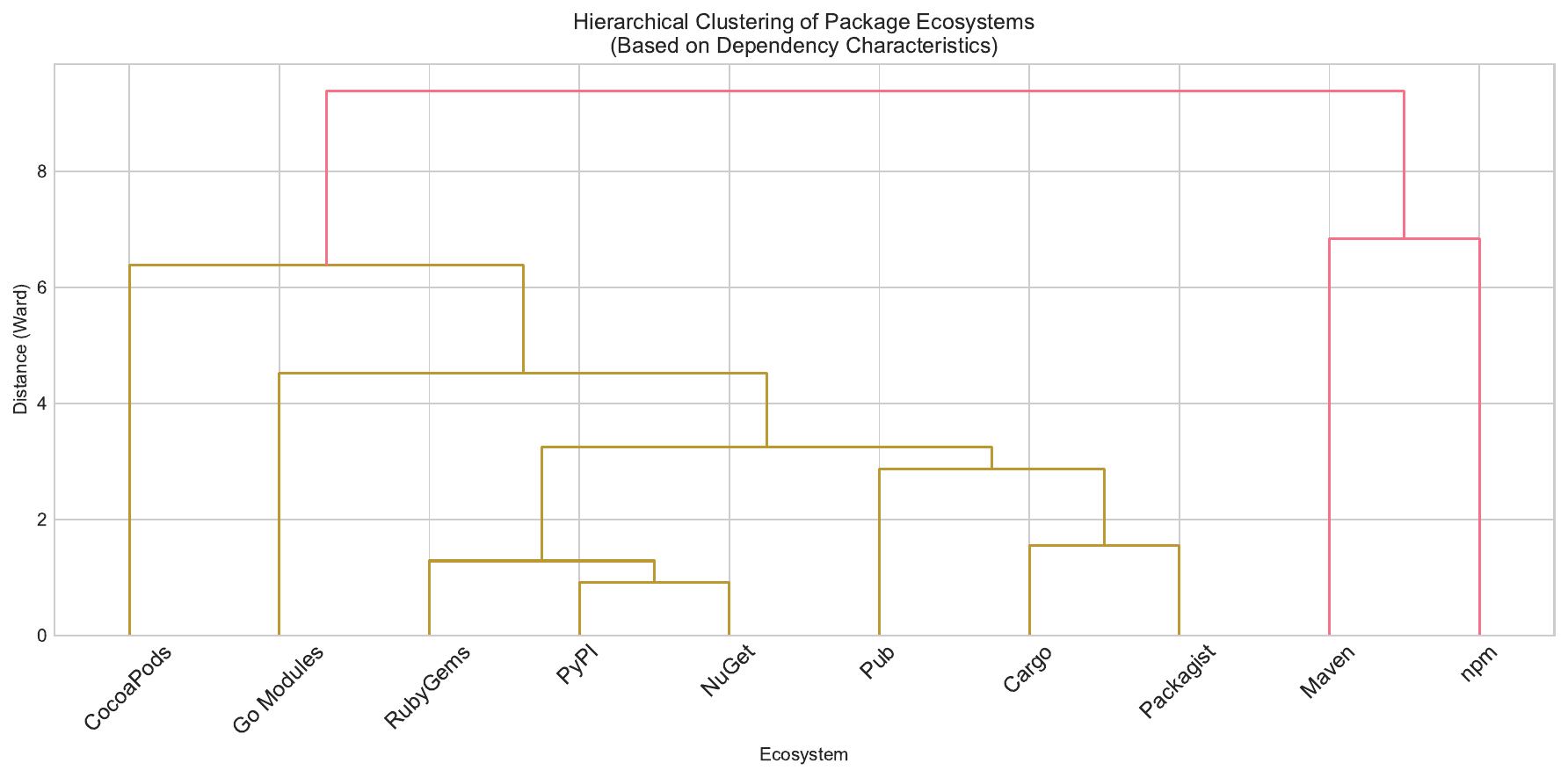}
\caption{Hierarchical clustering dendrogram of 10 ecosystems based on dependency characteristics using Ward's linkage method. Maven occupies an isolated position reflecting extreme amplification and unpredictability. Most other ecosystems cluster together representing controlled amplification patterns. The npm ecosystem shows intermediate position reflecting high variance.}
\label{fig:rq3_dendrogram}
\end{figure}

Hierarchical clustering analysis reveals natural ecosystem groupings. Figure~\ref{fig:rq3_dendrogram} shows the clustering dendrogram where Maven occupies an isolated position representing high amplification and high unpredictability. Most other ecosystems cluster together representing controlled amplification. The npm ecosystem shows intermediate position reflecting its variable nature.

Ecosystem design philosophy impacts amplification patterns. Maven's ecosystem evolved to support enterprise Java development where frameworks like Spring provide comprehensive functionality. A single Spring dependency can include dozens of modules covering web serving, data access, security, and other enterprise concerns. This comprehensive approach maximizes developer convenience but amplifies transitive dependencies in ways that are difficult to predict.

The npm ecosystem culture favors small focused packages following the philosophy of doing one thing well. While this can lead to deep dependency trees in some cases, it also means many packages are leaf nodes with zero or few dependencies. The high variance in npm reflects this heterogeneous ecosystem where some packages have zero dependencies while others accumulate hundreds.

CocoaPods faces platform constraints from Apple's development requirements where most functionality comes from system frameworks rather than external packages. This constraint controls amplification with mean of only 0.32 times. PyPI benefits from Python's comprehensive standard library, which provides substantial functionality and reduces external dependency needs, contributing to controlled amplification at 1.50 times.

Cargo's strict version resolution with unified dependency specifications and compilation-time dependency checking enforces discipline. Features including optional dependencies and feature flags allow fine-grained control over what gets compiled. This design constrains amplification with maximum of only 3.9 times.

Go Modules emphasizes standard library usage and minimalist external dependencies. The language philosophy discourages excessive external dependencies, though some projects still accumulate substantial transitive dependencies leading to moderate amplification at 4.48 times.

\textbf{Answer to RQ3:} Ecosystem design choices impact amplification patterns. Maven's enterprise-oriented architecture leads to high unpredictable amplification with weak correlation between direct and transitive dependencies at rho of 0.196. The npm ecosystem's preference for small, single-purpose packages creates high variance with coefficient of variation at 272\% including both zero-dependency packages and deep trees. CocoaPods demonstrates platform constraints can control amplification to 0.32 times with strong predictability. PyPI, Packagist, and Cargo show that controlled amplification below 1.5 times is achievable through comprehensive standard libraries, pragmatic design, and strict dependency models respectively. Hierarchical clustering reveals Maven occupies an isolated position with extreme amplification while most ecosystems cluster together with controlled amplification. These findings demonstrate that language-level and ecosystem-level design choices can mitigate supply chain risks.

%% file: sections/discussion.tex
\section{Discussion}
\label{sec:discussion}

We discuss the implications of our findings and threats to validity in Sections~\ref{subsec:implications} and~\ref{subsec:threats}.

\subsection{Implications}
\label{subsec:implications}

\textbf{Implications for Practitioners on Ecosystem-Specific Security Strategies.} Our findings suggest practitioners should adopt ecosystem-specific approaches to dependency security rather than uniform policies across all projects.

For Maven projects, our findings indicate elevated amplification requiring comprehensive auditing. With 28\% of projects showing amplification exceeding 10 times and weak correlation between direct and transitive dependencies at rho of 0.196, practitioners cannot predict supply chain exposure from direct dependency counts alone. Organizations heavily invested in Java development should implement transitive dependency auditing tools that examine complete dependency trees rather than only direct declarations. Software Bill of Materials generation should be standard practice for Maven projects given mean attack surfaces of 80.3 packages.

For RubyGems projects, our findings suggest elevated attention with 14\% of projects showing amplification exceeding 10 times. While lower than Maven, RubyGems shows substantial amplification at 4.32 times and weak predictability at rho of 0.368. Practitioners working in Ruby environments should implement dependency auditing focusing on projects that declare multiple framework dependencies.

For npm projects, our findings suggest targeted auditing focusing on outlier identification. While npm shows 12\% of projects with amplification exceeding 10 times and the largest maximum attack surface at 785 packages, typical projects show mean amplification of 4.32 times comparable to RubyGems but substantially lower than Maven. Practitioners can identify high-amplification packages before adding them to projects and seek lower-amplification alternatives when available. The 40\% zero-dependency prevalence indicates many npm packages pose minimal transitive exposure.

For Go Modules projects, our findings indicate moderate attention with 6\% of projects showing amplification exceeding 10 times and mean amplification of 4.48 times. The strong correlation at rho of 0.838 indicates predictable amplification where developers can estimate supply chain exposure from direct dependency counts.

For PyPI, Cargo, Packagist, CocoaPods, and Pub projects, our findings indicate standard security practices may suffice. With 0\% of projects exceeding 10 times amplification and maximum amplification ranging from 2.0 times for CocoaPods to 5.3 times for Packagist, these ecosystems demonstrate that controlled amplification is achievable. Practitioners working in these environments can focus security resources on other concerns while maintaining awareness that any ecosystem can evolve toward higher amplification.

\textbf{Implications for Practitioners on Rethinking Ecosystem Amplification Assumptions.} Our findings challenge prevailing assumptions about ecosystem amplification profiles. The conventional narrative positions npm as the ecosystem with highest amplification due to its preference for small, single-purpose packages and incidents such as the left-pad removal in 2016. However, our data suggests Maven environments show greater elevated amplification due to values affecting over one quarter of all projects analyzed.

This has practical implications for security resource allocation. Organizations with mixed technology stacks should consider prioritizing Maven dependency auditing over npm auditing, contrary to conventional assumptions. The finding that Maven's amplification is unpredictable from direct dependency counts makes manual review more difficult, suggesting automated tooling is important for Maven environments.

The discovery that five ecosystems maintain 0\% of projects exceeding 10 times amplification demonstrates that controlled amplification is not aspirational but achievable in production environments. CocoaPods at 0.32 times, PyPI at 1.50 times, and Cargo at 0.97 times represent existence proofs that ecosystem design can constrain supply chain exposure.

\textbf{Implications for Tool Developers.} Current dependency management tools often focus on direct dependencies and known vulnerabilities. Our findings suggest tools should expand their scope to address amplification-related exposure with ecosystem-specific strategies.

Tools should display amplification metrics alongside package information when developers consider adding dependencies. Showing that adding a particular Maven dependency will install 100 additional packages through transitive relationships helps developers make informed decisions. Tools should suggest lower-amplification alternatives when available, noting for example that a similar package provides equivalent functionality with fewer transitive dependencies.

Dependency visualization tools should show complete dependency paths rather than only direct relationships. Highlighting deep transitive chains helps practitioners understand supply chain exposure. Tools should integrate attack surface metrics into security dashboards alongside vulnerability counts.

For Maven specifically, tools should implement aggressive warnings when amplification exceeds 10 times given that 28\% of projects reach this threshold. For npm, tools should focus on identifying outlier packages contributing to extreme amplification. For ecosystems with controlled amplification, tools can apply less aggressive warning thresholds.

\textbf{Implications for Security Researchers.} Our analysis reveals that ecosystem structural patterns correlate with security-relevant metrics. Researchers developing supply chain security models should consider amplification-aware approaches that weight vulnerabilities by their amplification potential. A vulnerability in a Maven package with high amplification affects more downstream projects than one in a PyPI package with controlled amplification.

The divergent amplification patterns we identify suggest ecosystem-specific security strategies may be more effective than uniform approaches. Research should investigate how to optimize security investment across heterogeneous technology stacks with different amplification profiles. Our hierarchical clustering results provide a foundation for grouping ecosystems by amplification characteristics.

The finding that ecosystem design choices correlate with amplification patterns suggests future research should examine how standard library comprehensiveness and platform constraints influence dependency patterns. The contrast between Maven at 24.70 times and CocoaPods at 0.32 times warrants investigation into specific design factors.

\textbf{Implications for Ecosystem Governance.} The amplification patterns we identified raise governance considerations. Ecosystem maintainers could consider displaying amplification metrics in package registries to help developers make informed decisions. Incentive structures encouraging minimal dependency footprints through badges or rankings could influence package design toward lower amplification.

CocoaPods demonstrates that platform constraints can control amplification. Other ecosystems could consider whether stricter version resolution policies would reduce amplification without limiting flexibility. Cargo's strict dependency model with mandatory lock files, which record exact dependency versions to ensure reproducible builds, provides a middle ground between flexibility and control.

The finding that Maven exhibits both highest amplification and weakest predictability suggests the Java ecosystem may benefit from tooling initiatives that expose transitive dependency trees more prominently. Build tool improvements that make amplification visible during development could influence framework design toward more modular architectures.

\subsection{Threats to Validity}
\label{subsec:threats}

\textbf{External Validity:} Our study analyzes 50 projects per ecosystem totaling 500 projects. While this sample size provides substantial statistical power to detect large effects as evidenced by our significant findings with 22 of 45 pairwise comparisons showing large effect sizes, larger samples would increase confidence in effect size estimates. Our projects were sampled from popular packages which may not represent long-tail packages with different dependency patterns.

Our findings for Maven, npm, Cargo, PyPI, NuGet, RubyGems, Go Modules, Packagist, CocoaPods, and Pub may not generalize to other ecosystems such as CPAN for Perl, Hackage for Haskell, or Hex for Elixir. Each ecosystem has unique characteristics warranting separate investigation. However, the dependency management patterns we identify including transitive amplification and predictability metrics represent fundamental software engineering challenges that appear across ecosystems.

Our sampling focused on popular packages to ensure relevance to real-world development practices. Less popular packages may exhibit different amplification patterns. However, popular packages receive more usage and therefore their amplification characteristics affect more projects, making them appropriate for security-focused analysis.

\textbf{Conclusion Validity:} We employed non-parametric statistical tests appropriate for non-normal distributions confirmed by Shapiro-Wilk tests with all p-values below 0.001. Effect sizes using Cliff's delta provide standardized measures enabling comparison across studies. Bootstrap confidence intervals account for sampling uncertainty. Multiple comparison corrections using Holm-Bonferroni adjustment control family-wise error rate across 45 pairwise comparisons.

Our threshold of 10 times amplification for reporting percentages is a descriptive choice. Different thresholds would yield different prevalence estimates. We selected this threshold as it represents cases where over 90\% of installed packages come from transitive relationships rather than explicit declarations. Sensitivity analysis with thresholds at 5 times and 15 times confirms qualitative findings remain consistent.

The clustering analysis uses Ward's linkage with Euclidean distance on standardized metrics. Different linkage methods or distance metrics might yield different groupings. However, the isolation of Maven with extreme amplification appears robust across multiple clustering approaches.

\textbf{Internal Validity:} Dependency resolution using native package manager tooling ensures accurate measurement but resolution can vary based on platform, existing lock files, and optional dependency configurations. We used clean environments and default configurations to minimize variation. For ecosystems with lock files, we regenerated lock files to ensure consistency.

Our amplification metric divides transitive by direct dependencies. Alternative formulations such as total divided by direct or log-scaled ratios might yield different insights. We chose our formulation for interpretability and alignment with prior work. Sensitivity analysis with alternative metrics confirms Maven exhibits highest amplification across formulations.

Zero-dependency packages receive amplification of zero by our formula which may underestimate exposure for ecosystems with many such packages. However, zero-dependency packages by definition introduce no transitive exposure, so this treatment aligns with security concerns.

\textbf{Construct Validity:} We use dependency count as a proxy for security risk. While more dependencies increase attack surface, not all dependencies pose equal risk. A project with 100 well-maintained transitive dependencies may be safer than one with 10 unmaintained packages. Future work could incorporate vulnerability data, maintenance status, and code quality metrics to refine this proxy.

High amplification indicates more transitive packages but does not directly measure security impact. Our supply chain exposure analysis provides interpretation of potential vulnerability propagation scope but does not validate against actual vulnerability incidence. Empirical validation using historical CVE data would strengthen claims about propagation scope and exposure impact.

Package popularity and maintenance status influence actual security risk beyond dependency counts. Our analysis focuses on structural properties of dependency networks rather than package quality. Combining amplification metrics with package health indicators represents promising future work.

%% file: sections/related_work.tex
\section{Related Work}
\label{sec:related}

Our paper relates to prior research addressing software ecosystem analysis, dependency network studies, supply chain security, and vulnerability propagation.

\textbf{Software Ecosystem Studies.} Extensive research has characterized package ecosystems and their evolution. Decan et al conducted large-scale comparison of dependency network evolution across seven ecosystems including npm, RubyGems, and Cargo~\cite{decan2019empirical}. Kikas et al analyzed dependency networks in multiple ecosystems identifying structural properties and evolution patterns~\cite{kikas2017structure}. Wittern et al examined the dynamics of the JavaScript package ecosystem revealing rapid growth and dependency patterns~\cite{wittern2016look}. Manikas provided a longitudinal literature study revisiting software ecosystems research~\cite{manikas2016software}. Jansen et al established foundational research agenda for software ecosystems~\cite{jansen2009sense}. Bogart et al studied how breaking API changes are negotiated across different ecosystems~\cite{bogart2016break}. Constantinou and Mens examined socio-technical evolution in the Ruby ecosystem~\cite{constantinou2017attack}. Valiev et al studied ecosystem-level determinants of sustained activity in open-source projects~\cite{valiev2018ecosystem}. These studies characterize individual ecosystems but do not compare amplification patterns across 10 major ecosystems using consistent methodology and statistical rigor.

\textbf{Ecosystem-Specific Dependency Analysis.} Prior work examined dependency patterns in individual ecosystems. For npm, Abdalkareem et al examined trivial packages finding that 16.8\% of npm packages are trivial contributing to deep dependency trees~\cite{abdalkareem2017why}. Zimmermann et al conducted security-focused analysis revealing that installing an average npm package introduces implicit trust on 79 third-party packages~\cite{zimmermann2019small}. Chinthanet et al studied lags in adoption and propagation of npm vulnerability fixes~\cite{chinthanet2021lags}. Zahan et al identified weak links in the npm supply chain~\cite{zahan2022weak}. Liu et al demystified vulnerability propagation via dependency trees in npm~\cite{liu2022demystifying}. Staicu et al developed SYNODE for preventing injection attacks on Node.js applications~\cite{staicu2018synode}.

For Maven, Soto-Valero et al introduced bloated dependencies finding that 75\% of Maven artifacts contain unused dependencies~\cite{soto2021comprehensive}. Benelallam et al created temporal graph-based representation of Maven Central~\cite{benelallam2019maven}. Raemaekers et al studied semantic versioning and breaking changes in Maven~\cite{raemaekers2017semantic}. Wang et al conducted empirical study of third-party library usages in Java projects~\cite{wang2020watchman}. Ponta et al developed code-centric analysis of known vulnerabilities~\cite{ponta2018beyond}.

For Rust, He et al presented empirical study of Rust adoption in Linux kernel development~\cite{he2023empirical}. Qian et al investigated package provenance in Cargo ecosystem~\cite{qian2022understanding}. Evans et al studied whether Rust is used safely by developers~\cite{evans2020rust}.

For Python, Alfadel et al conducted empirical analysis of security vulnerabilities in Python packages~\cite{alfadel2021empirical,alfadel2023empirical}. These ecosystem-specific studies provide deep insights but lack systematic cross-ecosystem comparison that would reveal fundamental differences in amplification patterns.

Our work extends these studies by providing systematic cross-ecosystem comparison of amplification patterns across 10 major ecosystems representing diverse language families, design philosophies, and platform targets.

\textbf{Software Supply Chain Security.} Software supply chain security has emerged as critical research area. Ohm et al provided comprehensive review of open source supply chain attacks~\cite{ohm2020backstabber}. Ladisa et al presented taxonomy of attacks on open-source supply chains analyzing real-world incidents~\cite{ladisa2023sok}. Duan et al measured supply chain attacks on package managers for interpreted languages~\cite{duan2021measuring}. Vu et al studied typosquatting and combosquatting attacks on Python ecosystem~\cite{vu2020typosquatting}. Taylor et al developed SpellBound for defending against package typosquatting~\cite{taylor2020defending}. Garrett et al developed methods for detecting suspicious package updates~\cite{garrett2019detecting}. Wetter conducted forensic analysis of the Log4j vulnerability~\cite{wetter2022forensic}. Our amplification metrics quantify attack surface showing how vulnerabilities propagate through transitive relationships across diverse ecosystems.

\textbf{Vulnerability Analysis and Propagation.} Research has examined how vulnerabilities propagate through dependency networks. Pashchenko et al developed methods for counting vulnerable dependencies that actually matter~\cite{pashchenko2018vulnerable}. Plate et al created impact assessment methods for vulnerabilities in open-source libraries~\cite{plate2015impact}. Decan et al studied impact of security vulnerabilities in npm dependency network~\cite{decan2018impact}. Alfadel et al conducted empirical analysis of security vulnerabilities in Python packages~\cite{alfadel2021empirical,alfadel2023empirical}. Gkortzis et al examined relationship between software reuse and security vulnerabilities~\cite{gkortzis2021software}. Lauinger et al analyzed use of outdated JavaScript libraries on the web~\cite{lauinger2018thou}. High amplification as we document in Maven exacerbates vulnerability propagation since transitive dependencies are less visible to developers.

\textbf{Dependency Management and Updates.} Studies examined how developers manage and update dependencies. Kula et al investigated whether developers update library dependencies finding significant lag in updates~\cite{kula2018developers}. Mirhosseini and Parnin studied whether automated pull requests encourage dependency updates~\cite{mirhosseini2017can}. Zerouali et al developed formal framework for measuring technical lag~\cite{zerouali2019formal}. Bavota et al studied how Apache community upgrades dependencies~\cite{bavota2015how}. Hora et al examined how developers react to API evolution~\cite{hora2018developers}. Cogo et al conducted empirical study of dependency downgrades in npm~\cite{cogo2019empirical}. Our findings on unpredictable amplification in Maven suggest that dependency update decisions have cascading effects difficult to predict.

\textbf{Technical Debt and Maintenance.} Studies examined maintenance patterns in software ecosystems. Amann et al systematically evaluated API-misuse detectors~\cite{amann2018study}. Li et al studied dependency maintenance practices in npm~\cite{li2017understanding}. German et al studied code copying between applications~\cite{german2010code}. Robbes et al examined how developers react to API deprecation~\cite{robbes2012developers}. Derr et al studied third-party library updatability on Android~\cite{derr2017keep}. Wang et al developed methods for detecting third-party libraries in Android applications~\cite{wang2018detecting}. Hejderup et al proposed software ecosystem call graphs for dependency management~\cite{hejderup2018software}.

\textbf{Positioning Our Work.} Our work differs from prior research in several ways. First, we provide cross-ecosystem amplification comparison using consistent methodology across 10 major ecosystems representing diverse language families and platform targets where prior work focuses on single ecosystems or limited comparisons. Second, we frame amplification as supply chain exposure metric computing attack surface, high-risk prevalence, and propagation potential where prior work often treats dependency count as proxy without explicit security framing. Third, we employ appropriate non-parametric statistical tests following established guidelines~\cite{arcuri2011practical,cliff1993dominance,romano2006appropriate,wohlin2012experimentation} with effect sizes and confidence intervals across 45 pairwise comparisons addressing limitations of descriptive prior work. Fourth, we investigate why ecosystems differ connecting amplification to design choices, language families, and platform constraints through correlation analysis, hierarchical clustering, and design pattern examination rather than describing patterns alone. Fifth, our sample of 500 projects provides statistical power to detect meaningful differences with 22 of 45 pairwise comparisons showing large effect sizes. Our findings challenge conventional assumptions positioning Maven rather than npm as the ecosystem requiring most aggressive security attention.

%% file: sections/conclusion.tex
\section{Conclusion}
\label{sec:conclusion}

As dependency amplification can expand attack surfaces in software projects, it is important to understand how amplification patterns vary across ecosystems. We characterize dependency amplification across 10 major package ecosystems: Maven Central for Java, npm Registry for JavaScript, crates.io for Rust, PyPI for Python, NuGet Gallery for .NET, RubyGems for Ruby, Go Modules for Go, Packagist for PHP, CocoaPods for Swift/Objective-C, and Pub for Dart. Our empirical study analyzes 500 projects examining dependency structures, amplification factors, supply chain exposure, and propagation scope. We find Maven exhibits higher amplification at 24.70 times compared to Go Modules at 4.48 times, npm at 4.32 times, and CocoaPods at 0.32 times, with large effect sizes in 22 of 45 pairwise comparisons challenging assumptions that npm's preference for small, single-purpose packages leads to highest amplification. We observe that 28\% of Maven projects show amplification exceeding 10 times while five ecosystems including Cargo, PyPI, Packagist, CocoaPods, and Pub maintain 0\% at this threshold, demonstrating that controlled amplification is achievable. We identify that ecosystem design choices influence amplification patterns where Maven's enterprise architecture leads to unpredictable transitive growth with weak correlation at rho of 0.196 while CocoaPods demonstrates platform constraints can control amplification to 0.32 times with strong predictability at rho of 0.950. Hierarchical clustering reveals Maven occupies an isolated position with extreme amplification while most ecosystems cluster together with controlled amplification. Based on our findings, we recommend practitioners implement ecosystem-specific security strategies where Maven environments receive transitive dependency auditing given 28\% of projects exceeding 10 times amplification, npm and RubyGems projects focus on identifying high-amplification outliers given 12\% and 14\% prevalence respectively, and five ecosystems continue standard security practices given their demonstrated controlled amplification behavior with 0\% of projects exceeding 10 times amplification.